\documentclass[twocolumn,showpacs,preprintnumbers,amsmath,amssymb,pre]{revtex4}
\usepackage{times}
\usepackage{graphicx}
\usepackage{appendix}
\usepackage{bm}

\newcommand{\f}{\frac}

\newcommand{\nn}{\nonumber}
\newcommand{\bea}{\begin{eqnarray}}
\newcommand{\eea}{\end{eqnarray}}

\def\d{\delta}

\def\nn{\nonumber}
\def\ni{\noindent}

\def\la{\langle}
\def\ra{\rangle}
\def\e{\epsilon}
\def\o{\omega}

\def\g{\gamma}

\def\bx{{\bf x}}

\def\by{{\bf y}}

\begin{document}
\title{Time auto-correlation function and Green-kubo formula: A study on disordered harmonic chain}
\author{Anupam Kundu$^1$}
\affiliation{$^1$ Raman Research Institute, Bangalore 560080, India}
\begin{abstract}

We have considered heat conduction in a one-dimensional mass disordered harmonic chain of $N$ particles connected to two Langevin type reservoirs at different temperatures. An exact expression for the boundary 
heat current-current auto-correlation function in the 
non-equilibrium steady state (NESS) is obtained in terms of non-equilibrium phonon 
Green's functions. The time integral of the 
correlation function gives expected result, both in non-equilibrium as well as 
equilibrium cases. Using the form of this correlation function we show that asymptotic system size dependence of current fluctuation in NESS for a mass disordered harmonic chain is $N^{-\alpha}$ for different boundary conditions. For free and fixed boundary conditions we get $\alpha=1/2$ and $3/2$ respectively, while for pinned case the fluctuation decays exponentially with system size.  
\end{abstract}

\pacs{}
\maketitle
\section{ I\lowercase{ntroduction}}\label{intro}
Time correlation functions are useful quantities in the study of transport processes. They are related to various transport 
coefficients. For example, the diffusion constant of a Brownian particle is given by the integral of the equilibrium 
velocity-velocity 
time auto-correlation function. Similarly the friction coefficient of an over-damped particle is also related to the time 
correlation function of the instantaneous force experienced by the particle. 
Let us consider a stochastic process described 
by the vector $\bx(t)$. 
Then the time correlation function of any quantity $A(t)=A(\bx(t))$ is defined as: $\la A(t)A(t')\ra$ where $\la \ra$ 
represents the average over initial conditions and trajectories. 
In terms of phase space variables, $\la A(t)A(t')\ra$ is given by 
\bea
\la A(t)A(t')\ra = \int d\bx \int d\by A(\bx)A(\by)W(\bx,t;\by,t')P(\by,t')
\eea
where, $P(\by,t')$ is the probability of $\by$ at time $t'$ and $W$ is the transition probability from $\by$ to $\bx$ in time 
$t-t'$.
In general the equilibrium time 
correlation function of some quantity is related to the response of a system to small perturbations. These relations are called Green-Kubo formula (GK) \cite{green54,kubo57b}. 
 
For the case of heat transport the GK formula relates the response of a system to a small temperature gradient to the equilibrium heat current auto correlation function. The response to temperature gradient defines the thermal conductivity $\kappa$ and the GK formula gives:
\begin{eqnarray}
\kappa &=& \lim_{\tau\rightarrow\infty}\lim_{L\rightarrow\infty} 
\frac{1}{k_BT^2L^d}
\int_0^\tau dt \langle J(t) J(0)\rangle ~,
\label{kubo_n}
\end{eqnarray}
where $J(t)$ is the heat current through the system at time $t$ and $L$ is the linear dimension of a $d$-dimensional system.
In Eq.~(\ref{kubo_n}) the order of the limits is very important. 
Although this is a very useful formula, there are some difficulties associated with this formula.
The formula in Eq.~(\ref{kubo_n}) is not applicable to small mesoscopic structures. 
Also in case of anomalous transport, which occurs in many low dimensional systems, the thermal conductivity diverges 
\cite{lepri03,dharrev}. In such cases it is not possible to take the limits as in Eq.~(\ref{kubo_n}). 
There are various derivations of this formula \cite{others,lutt}.   
Recently we have derived a formula similar to Eq.~(\ref{kubo_n}) for open systems, which is applicable to systems of arbitrary size in any dimensions \cite{kundu}. This derivation uses Fokker-planck description of stochastic systems 
and hence is only applicable for those currents, which can be expressed in terms of phase space variables 
(\emph{e.g.} currents inside the bulk of the system). 
Since boundary currents naturally involve noises explicitly, 
derivation given in \cite{kundu} is not be applicable for them. 
General expectation is, for boundary currents also one can proof a open finite system GK formula as given in \cite{kundu}. 
In this paper we explicitly calculate boundary current-current auto correlation function in 
the context of heat transport 
for a finite mass disordered harmonic chain in NESS and show that integration of the equilibrium correlation function 
gives the NESS current.

There are few examples where exact time auto-correlation functions in equilibrium state have been obtained for 
many-particle systems. For Hamiltonian systems some examples of exact calculations are velocity auto-correlation function 
for ordered harmonic lattices \cite{Mazur} and for a one dimensional gas of elastically colliding hard rods \cite{Jepsen}. Recently authors of \cite{morgado} have shown explicitly that integration of the heat current auto-correlation function gives the current in non-equilibrium steady state for a two particle harmonic system. 
In this paper we obtain an exact expression for the time auto-correlation function for boundary  
heat current in the NESS for  mass disordered harmonic chains of arbitrary length, 
expressed in terms of the 
non-equilibrium Green's functions. We show that it satisfies the GK formula derived in \cite{kundu}. Using this correlation function we also calculate the asymptotic system size scaling of fluctuations in current in NESS.

The paper is organised as follows. In sec.~({\ref{model}}) we give the description of the model, 
define some relevant quantities and notations and calculate the current in the NESS.  
In sec.~(\ref{corr}) we present the calculation of the time correlation function.  
In sec.~({\ref{discsn}}) we discuss our results and finally in sec.({\ref{conclu}}) we conclude.

\noindent
\section{ D\lowercase{efinition of model}}\label{model}
We consider a chain of oscillators of N particles 
described by the Hamiltonian $H$ :

\begin{eqnarray}
H=\sum_{l=1}^N [\f{1}{2}m_l \dot{x}_l^2 &+&\f{1}{2}k_o{x}_l^2]+\sum_{l=1}^{N-1}\f{1}{2}k(x_{l+1}-x_l)^2  \nonumber \\ 
&+& \f{1}{2}k'(x_1^2+x_N^2) ~,               
\label{hamilto}              
\end{eqnarray} 
where {${x}_l$} are displacements of the particles about their equilibrium positions, $k$, $k_0$ are the inter-particle 
and on-site spring constants respectively, and $m_l$ is mass of the $l^{th}$ particle. $k'$ is the spring constant of the potentials at the boundaries. For different values of $k'$ and $k_0$ we get different boundary-conditions (BCs). If $k'$ and $k_0$ both are zero we get free BC, otherwise we get fixed BC ($k' \neq 0$ and $k_0 = 0$) and pinned case ($k_0 \neq 0$). 
The particles $1$ and $N$ are connected to 
two white noise heat baths of temperatures $T_L$ and $T_R$ respectively. 
The equation of motion of the $l^{th}$ particle is given by \cite{dhar01}
\bea
m_l\ddot{x}_l &=& -k(2x_l - x_{l-1} - x_{l+1}) - k_o x_l  \nn \\
&&-\d_{l,1} [ (k'-k)x_l +{\gamma_L}{\dot{x}}_1 - {\eta}_L] \nn \\
&&-\d_{l,N} [ (k'-k)x_l +{\gamma_R}\dot{x}_N - {\eta}_R~] \nn  \\
&&~~~{\rm{where}}~~~~~l=1,2...N~~{\rm{and}}~~x_0=x_{N+1}=0
\label{eqmlang}              
\eea
where ${\eta}_{L,R}(t)$  are Gaussian noise terms with zero mean and related to the dissipative terms with these relations
\bea   
\la {\eta}_{L,R}(t){\eta}_{L,R}(t')\ra &=& 2{\gamma}_{L,R}{T}_{L,R}\delta(t-t') \nn \\
\la {\eta}_{L}(t){\eta}_{R}(t')\ra&=&0,~~~~\la {\eta}_{L,R}(t)\ra=0
\label{fl-disi}
\eea
(In this paper we have set $K_B=1$.)
To define the local energy current inside the chain we first define the
local energy density 
associated with the $l^{\rm th}$ particle (or energy at the lattice site $l$) as follows:
\bea
\e_1 &=& \f{p_1^2}{2 m_1} + \f{k_ox_1^2}{2} + \f{k'x_1^2}{2}+ \f{k}{4}(x_1-x_2)^2 ~, \nn \\
\e_l &=& \f{p_l^2}{2 m_l}+ \f{k_ox_l^2}{2} + \f{k}{4}
  [~(x_{l-1}-x_l)^2+ (x_l-x_{l+1})^2~]~, \nn \\
&&~~~~~~~~~~~~~~~~~~{\rm for }~~l=2,3...N-1 \nn \\
 \e_N &=& \f{p_N^2}{2 m_N} +\f{k_ox_N^2}{2}+\f{k'x_N^2}{2}+ \f{k}{4} (x_{N-1}-x_N)^2~. \label{enerdef1} 
\eea
Using this energy density we write a continuity equation, from which we 
get two instantaneous currents $j_{L}$ and $j_{R}$ which are flowing from the left and right reservoirs into the system
respectively. These currents are given by \cite{lepri03,dharrev} 
\bea
{j}_{L}(t)&=& - {\gamma}_L {\dot{x}}_1^2(t)+ {\eta}_L(t){\dot{x}}_1(t)~, \nn \\ 
{\rm and}~~{j}_{R}(t)&=&- {\gamma}_R{\dot{x}}_N^2(t)+  {\eta}_R (t)\dot{x}_N(t)~.
\label{curdef}
\eea
  
\noindent
In order to obtain the steady state properties we have to find out the steady state solution of the Eq.~(\ref{eqmlang}). For that we write  Eq.~(\ref{eqmlang}) in Matrix form as:
\begin{eqnarray}
M\ddot{X} + \Gamma \dot{X} + \Phi X = {\bf{\eta}}(t),
\label{eqmotmat}              
\end{eqnarray}
where,$X,\eta$ are column vectors with elements ${[X]}^T = ({x}_1,{x}_2,....{x}_N)$,
${[{\bf{\eta}}]}^T = ({\eta}_L,0,....0,{\eta}_R)$ and $\Gamma$ is a $N \times N$ matrix with only 
non-vanishing elements ${[\Gamma]}_{11}={\gamma}_L$, ${[\Gamma]}_{NN}={\gamma}_R$. ${[\Phi]}_{N \times N}$ represents 
a tridiagonal matrix with elements \cite{droy-ordrd}
\bea
\Phi_{lm} &=& (k+k'+{k}_o){\delta}_{l,m} -k{\delta}_{l,m-1}~~{\rm{for}}~l=1 \nn \\
&=& - k{\delta}_{l,m-1} + (2k+{k}_o){\delta}_{l,m} -k{\delta}_{l,m+1}  \nn \\
&&~~~~~~~{\rm{for}}~~ 2 \le l \le N-1 \nn \\
&=& (k+k'+{k}_o){\delta}_{l,j} -k{\delta}_{l,m+1}~~{\rm{for}}~l = N~,
\eea
and $M_{lm}=m_l\delta_{lm}$ where $m_l$ is chosen uniformly from the range 
$[1-\Delta, 1+\Delta]$. 
If $\mathcal{G}^+(t)$ denotes the Green's function of the entire system then $\mathcal{G}^+(t)$ satisfies 
\bea
M\ddot{\mathcal{G}}^+(t) + \Gamma \dot{\mathcal{G}}^+(t) + \Phi \mathcal{G}^+(t) = \delta(t) I~,
\label{Grn-eom1}
\eea
It is easy to verify that $\mathcal{G}^+(t)={G}(t)\Theta(t)$ where ${G}(t)$ satisfies the  homogeneous 
equation 
\bea
M\ddot{G} + \Gamma \dot{G} + \Phi G = 0~,
\label{Geqm}
\eea 
with the initial conditions ${G}(0)=0$, 
$\dot{{G}}(0)= M^{-1}$. Here $\Theta(t)$ is the Heaviside function.
Assuming that the heat baths have been switched on at $t=-\infty$ we write the steady state solution of  
Eq.~(\ref{eqmotmat}) as
\bea
X(t) = \int_{-\infty}^{t} dt' G(t-t')\eta (t').
\label{soln}
\eea
 For equilibration we require that $G(t)\rightarrow 0$ as $t\rightarrow \infty$. 
From Eq.({\ref{soln}}), we get 
\bea
\dot{x}_1(t) &=& \int_{-\infty}^{t} dt_1 \Big[\dot{G}_{11}(t-t_1)\eta_L(t_1) \nn \\
&&+ \dot{G}_{1N}(t-t_1)\eta_R(t_1)\Big]~. 
\label{x1dot}
\eea
Next we calculate $\la j_{L}\ra$ in the NESS. Here $\la ...\ra$ denotes the average over the 
noise variables $\eta_L(t)$ and $\eta_R(t)$. From now we denote $\la j_{L}\ra$ by $j$. 
Putting $\dot{x}_1(t)$ from Eq.~(\ref{x1dot}) in 
the expression of $j_L(t)$ in Eq.~(\ref{curdef}) and using the noise correlation in Eq.~(\ref{fl-disi}) we get :
\bea
j&=& 
-\g_L\int_{-\infty}^{t}dt_1 \int_{-\infty}^{t}dt_2 
\Big[\dot{G}_{11}(t-t_1){\dot{G}}_{11}(t-t_2) \nn \\
&& \times \langle \eta_L(t_1)\eta_L(t_2)\rangle 
+ \dot{G}_{1N}(t-t_1)\dot{G}_{1N}(t-t_2) \nn \\ 
&& \times \langle \eta_R(t_1)\eta_R(t_2)\rangle \Big] 
+\int_{-\infty}^{t}dt_1 \dot{G}_{11}(t-t_1)\langle \eta_L(t)\eta_L(t_1)\rangle \nn \\
&&=2\g_L \left[\f{T_L}{2}\dot{G}_{11}(0) - (\g_LT_L A_1(0)+\g_RT_R A_N(0))\right], \nn \\
\label{jav0}  
\eea
where we have used the definition
\bea
A_i(t) = \int_{0}^{\infty}dt' \dot{G}_{1i}(t+t')\dot{G}_{1i}(t')~~~~~~\forall~~~t~.
\label{Ai}
\eea
We now note the following identity (for proof see Appendix A) 
\bea
\gamma_L A_1(t)+\gamma_R A_N(t)= \f{\dot{G}_{11}(t)}{2},
\label{identity}
\eea
which can be obtained from Eqs.(\ref{Ai},\ref{Geqm}). Using this in Eq.~(\ref{jav0}) we get
\bea 
j=2\g_L\g_R(T_L-T_R)A_N(0)~.
\label{Jav}
\eea
If we go to the frequency $\o$ space using the following definition 
\bea
G^+(\o)=\int_{0}^{\infty}dt~G(t)e^{i\o t},
\eea
we can identify that 
\bea
A_i(t)=\f{1}{2\pi}\int_{-\infty}^{\infty}\o^2 |G_{1i}^+(\o)|^2 e^{i\omega t},
\label{At}
\eea
and
\bea
G^+(\o)=\Big[-M\o^2+i\o\Gamma+\Phi \Big]^{-1}. 
\eea

With this identification we see that the 
expression given in Eq.~(\ref{Jav}) reduces to the form 
\bea
j=\f{(T_L-T_R)}{2\pi}\int_0^{\infty}~d\o~\mathcal{T}(\o),
\eea
where 
\bea
\mathcal{T}(\o)=4 \g_L \g_R~\o^2 |G_{1N}^+(\o)|^2,
\label{To}
\eea
 is the transmission coefficient for frequency $\o$. The above expression for the current $j$
is seen to be identical to the 
well-known expression for the current given in \cite{lebo-cash,dharroy}. 

\noindent
In the next section we proceed to obtain the time auto-correlation function $C_{\Delta T}(t,t')$ defined as: 
\bea
C_{\Delta T}(t,t')=\la j_{L}(t)j_{L}(t')\ra - \la j_L \ra^2,
\eea 
in the NESS. The subscript $\Delta T$ represents the difference between the temperature at the two ends {\emph{i.e.}} $\Delta T = T_L - T_R$. 
In the stationary state 
$ \la j_{L}(t)j_{L}(t')\ra $ will be a function of $|t-t'|$ only. Hence we set $t'=0$.
If we take $\Delta T =0$ in the expression of $C_{\Delta T}(t)$ we get the equilibrium auto-correlation which is denoted by 
$C_{0}(t)$ and we show that 
integral of $C_{0}(t)$ is related to the average current $\la j_{L}\ra$, whereas integral of $C_{\Delta T}(t)$ is related to its fluctuations in the NESS.

\noindent
\section{ C\lowercase{alculation of auto-correlation function}}\label{corr}
Using the forms of $j_L$ from Eq.~(\ref{curdef}) we write current current auto-correlation $\la j_L(t)j_L(0)\ra$ as: 
\bea
\langle j_L(t)j_L(0)\rangle&=&J_{L1}+J_{L2}+J_{L2}+J_{L4}~, \nn \\
{\rm where}&&~~~~\nn \\
J_{L1}&=&\gamma_L^2\langle \dot{x}_1^2(t)\dot{x}_1^2(0)\rangle, \nn \\
J_{L2}&=&-\gamma_L\langle \eta_L(t)\dot{x}_1(t)\dot{x}_1^2(0)\rangle, \nn \\
J_{L3}&=&-\gamma_L\langle \eta_L(0)\dot{x}_1^2(0)\dot{x}_1(t)\rangle, \nn \\
J_{L4}&=&\langle \eta_L(t)\dot{x}_1(t)\eta_L(0)\dot{x}_1(0)\rangle, 
\label{cc-cor}
\eea
where $t>0$.

\noindent
Now we will calculate all these $J$'s using  Eq.~(\ref{x1dot}) and Eq.~(\ref{fl-disi}). We will present the calculation 
of $J_{L1}$ explicitly and state the results for other $J$'s. 
Putting the form of $x_1(t)$ in the expression of $J_{L1}$ in Eq.~(\ref{cc-cor}) we get
\bea
J_{L1}=\gamma_L^2\int_{-\infty}^{t}dt_1\int_{-\infty}^{t}dt_2\int_{-\infty}^{0}dt_3\int_{-\infty}^{0}dt_4 \nn \\
~~~~~~\times K_1(t_1,t_2,t_3,t_4,t),
\label{J-L1}
\eea
Where $K_1(t_1,t_2,t_3,t_4,t)$ is given by
\bea
&&K_1(t_1,t_2,t_3,t_4,t)\nn \\
&&=\Big\la[\dot{G}_{11}(t-t_1)\eta_L(t_1)+\dot{G}_{1N}(t-t_1)\eta_R(t_1)] \nn \\
&&\times[\dot{G}_{11}(t-t_2)\eta_L(t_2)+\dot{G}_{1N}(t-t_2)\eta_R(t_2)] \nn \\
&&\times[\dot{G}_{11}(-t_3)\eta_L(t_3)+\dot{G}_{1N}(-t_3)\eta_R(t_3)] \nn \\
&&\times[\dot{G}_{11}(-t_4)\eta_L(t_4)+\dot{G}_{1N}(-t_4)\eta_R(t_4)]\Big\ra~.
\eea

\begin{figure}[t]
\includegraphics[width=9cm,height=5cm]{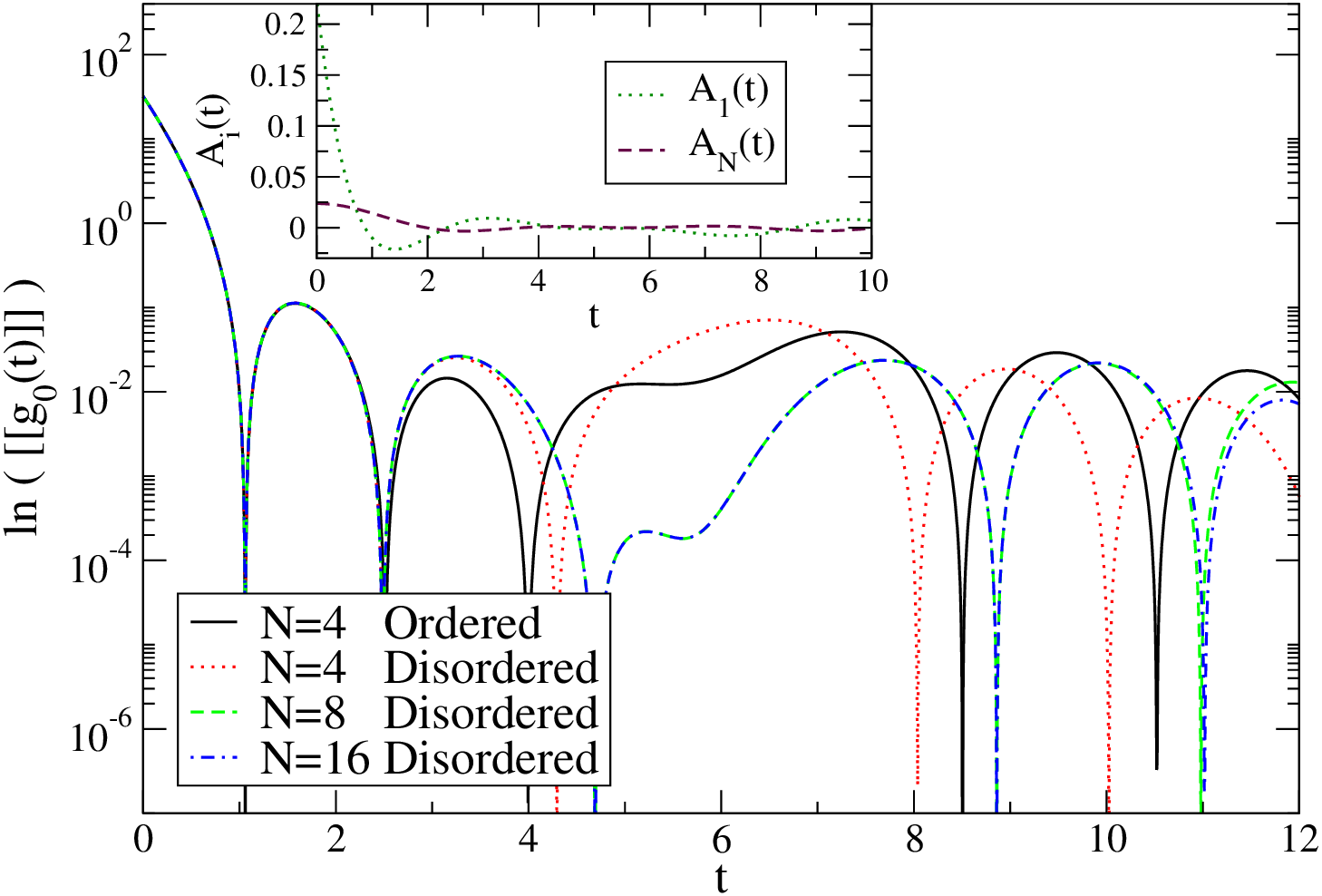}
\vspace{0.1cm}
\caption{(Color online) Plots of $[[g_0(t)]]$ vs. $t$ for $N=4$ and $N=8$. The parameters for the figure are $T_L=2.0, T_R=2.0$, $k=1.0$, $k_0=0.0$, $k'=0.0$, $\g_L=\g_R=2.5$ and $\Delta = 0.4$. Here $[[g_0(t)]]$ denotes disorder averaged $g_0(t)$. The average is done over $100$ disorder realisations. Inset shows the plots of $A_1(t)$ and $A_N(t)$ for $N=8$ for a single disorder configuration.
}
\label{fig-1}
\end{figure}
 
\ni
After taking the average over noises and using their Gaussian property, we get
\bea
&&K_1(t_1,t_2,t_3,t_4,t) \nn \\
&&=4(K_1^{(1)}(t_1,t_2,t_3,t_4,t) \d(t_1-t_2)\d(t_3-t_4) \nn \\
&&+K_1^{(2)}(t_1,t_2,t_3,t_4,t) \d(t_1-t_3)\d(t_2-t_4)\nn \\
&&+K_1^{(3)}(t_1,t_2,t_3,t_4,t) \d(t_1-t_4)\d(t_2-t_3))
\eea


\ni
where expressions for these $K_1^{'s}$ are given in Appendix B.

\ni 
Putting the expression of $K_1(t_1,t_2,t_3,t_4,t)$ in Eq.~(\ref{J-L1}) and arranging the terms we finally get
\bea
J_{L1} &=& 4\g_L^2 \big[{\big\{\g_LT_L A_1(0) + \g_R T_R A_N(0)\big\}}^2 \nn \\ 
&+& 2{\big\{\g_LT_L A_1(t) + \g_R T_R A_N(t)\big\}}^2\big], 
\label{J-L1-1}
\eea
where we have used the definitions of $A_i(t)$ in Eq.~(\ref{Ai}).~
\noindent
Similarly we calculate other $J$'s and their expressions are  
\bea
J_{L2} &=& -4\g_L^2T_L  [\f{1}{2}\dot{G}_{11}(0)\{\g_LT_LA_1(0) + \g_RT_RA_N(0)\},   \nn \\
J_{L3} &=& -4\g_L^2T_L  [\f{1}{2}\dot{G}_{11}(0)\{\g_LT_LA_1(0) + \g_RT_RA_N(0)\} \nn \\
&+& 2 \dot{G}_{11}(t)\{\g_LT_LA_1(t) + \g_RT_RA_N(t)\}], \nn \\
J_{L4} &=& 4\g_LT_L[\delta(t)\{\g_LT_LA_1(t) + \g_RT_RA_N(t)\} \nn \\
&+&\g_LT_L\{\f{1}{4}\dot{G}_{11}^2(0)\} ]~.\nn \\
\label{Jforms}
\eea
Collecting all the expressions for $J$'s from Eqs.~(\ref{J-L1-1}) and (\ref{Jforms}) in Eq.~(\ref{cc-cor})  
and subtracting ${\la j_L\ra}^2$ we finally obtain
\bea
C_{\Delta T}(t)&=&4\g_L T_L\big\{\g_L T_L A_1(0)+\g_R T_R A_N(0)\big\}\delta(t) \nn \\
&-&8{\g_L}^2\big[{\big\{\g_L T_L A_1(t)+\g_R T_R A_N(t)\big\}} \nn \\
&\times&{\big\{T_L\g_LA_1(t)+(2T_L-T_R)\g_RA_N(t)\big\}}\big]~,\nn \\ 
&=&4\g_L T_L\big\{\g_L T_L A_1(0)+\g_R T_R A_N(0)\big\}\delta(t) \nn \\ 
&& ~~~~~~-g_{\Delta T}(t)
\label{C_TLTR} 
\eea
where 
\bea
g_{\Delta T}(t) &=& 8{\g_L}^2\big[{\big\{\g_L T_L A_1(t)+\g_R T_R A_N(t)\big\}} \nn \\
&\times&{\big\{T_L\g_LA_1(t)+(2T_L-T_R)\g_RA_N(t)\big\}}\big]~, \label{gt}
\eea
 and we have used the identity in Eq.~(\ref{identity}). From the above expression of $g_{\Delta T}(t)$ we note that 
$g_{0}(t)$ is always positive.
Thus we have obtained a closed form expression for the non-equilibrium current-current 
auto-correlation function expressed in terms of the 
Green's function for a disordered harmonic chain of length $N$. The delta function appearing in the above equation is purely due to the white nature of the noises. More generally one can define the current operator on any bond on the harmonic chain. However the detailed form of the bond-correlation function is quite different from that of the boundary-correlation function. The notable difference that we find is the absence of the $\delta$-function peak.  We have verified that the integral of bond-correlation agrees with the value for the boundary-correlation.

\begin{figure}
\includegraphics[width=9cm,height=5cm]{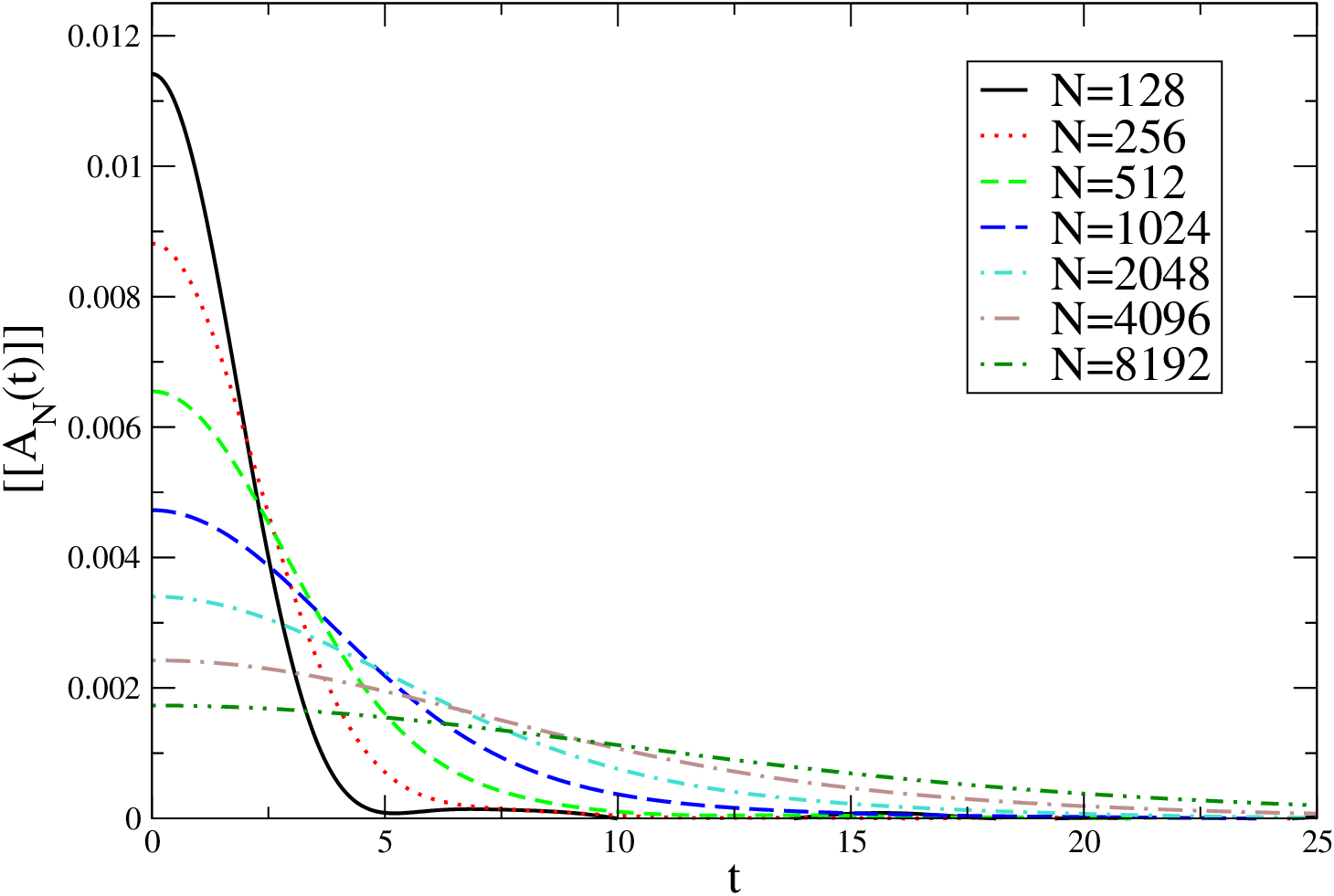}
\vspace{0.1cm}
\caption{(Color online) Plots of $[[A_N(t)]]$ vs. $t$ for different system sizes. The parameters for the figure are same as those for Fig.~\ref{fig-1}. $\Delta = 0.4$. 
}
\label{fig-2}
\end{figure}

\section{D\lowercase{iscussions}} \label{discsn}
In this section we plot the function $g_{\Delta T}(t) = 4\g_L T_L\big\{\g_L T_L A_1(0)+\g_R T_R A_N(0)\big\}\delta(t) -C_{\Delta T}$. To find the functional form of $g_{\Delta T}(t)$ we need to know the functional forms of the functions $A_i(t)$.  These functions can be obtained by Fourier transforming $\omega^2 |G_{1i}(\o)|^2$ as shown in Eq.~(\ref{At}). For a general N-particle mass disordered chain it is difficult to find analytical expressions for the functions $|G_{ij}^+(\o)|^2$.  For the ordered case $G_{ij}^+(\o)$ can be 
obtained analytically using the tridiagonal nature of the force matrix $\Phi$ (see for example Refn.~\cite{droy-ordrd}). 
 However in case of disordered chain, $G^+_{1N}(\o)$ and $G^+_{11}(\o)$ can be obtained through transfer matrix approach in which $G^+_{1N}(\o)$ and $G^+_{11}(\o)$ are expressed in terms of a product of $N$ random matrices \cite{dhar01}.  We numerically evaluate $G^+_{1N}(\o)$ and $G^+_{11}(\o)$ using this transfer matrix approach. 
We observe that at large $\o > \o_d=\f{km}{N\sigma^2}$, $\big[\big[|G^+_{1N}(\o)|^2\big]\big]$ decays as $e^{-aN\o^2}$ ($a$ is a positive constant) where $m=[[m_l]]$ and $\sigma^2=[[(m_l-m)^2]]$. Here $[[...]]$ denotes disorder average. This behaviour was proved analytically by Matsuda and Ishi \cite{MI} and was first observed numerically by Dhar \cite{dhar01}. 
Another observation made by Dhar was that for $\o < \o_d$ disordered average of $|G^+_{1N}(\o)|^2$ is almost identical to that of an ordered chain for both the BCs. We make use of this observation in this paper. Another observation which we made is that for $\o > \o_m$ the function $|G^+_{11}(\o)|^2$ decays as $1/\o^4$, where $\o_m$ is the maximum normal mode frequency. This $1/\o^4$ behaviour can be easily obtained through the transfer matrix approach. For small frequencies disorder average of $|G^+_{11}(\o)|^2$ oscillates with $\o$ and is again identical to that of ordered chain.
 
After integrating Eq.~(\ref{At}) numerically, we obtain $A_i(t)$ and $G_{ij}(t)$ and hence $g_0(t)$ for different system sizes with different disorder configurations. In Fig.~\ref{fig-1} we plot $[[g_{\Delta T}(t)]]$ versus $t$ for system sizes $N=4,~8$ and $16$ with free BC. We observe that the correlation functions for two system sizes remains almost identical at short times and starts being different significantly after some time scale. These observations 
can be made by looking at the dominant contributions of $\omega^2 |G^+_{1i}(\omega)|^2$ in the integrand of Eq.~(\ref{At}) for fixed $t$. At large $\o$ the functions $|G^+_{1N}(\o)|^2$ decays as $e^{-aN\o^2}$ ($a$ is a positive constant)\cite{MI,dhar01} whereas $|G^+_{11}(\o)|^2$ decays as $1/\o^4$. At small frequencies both $G^+_{1N}(\o)$ and $G^+_{11}(\o)$ are oscillating function of $\omega$ and the frequency of oscillation increases with system size $N$. As a result $A_1(t)$ is independent 
of system size $N$ at small times and starts depending on $N$ after some time scale, where contribution from small $\omega$ becomes important. Whereas, in case of $A_N(t)$, only a small range of $\omega$ contribute in the Fourier transform of $\omega^2|G^+_{1N}(\omega)|^2$(Eq.~(\ref{At})). For large $N$, at small times $A_1(t)$ is much larger that $A_N(t)$ and contributes most in $g_0(t)$, which makes $g_0(t)$ to be independent of $N$ at small times. Inset in Fig.~\ref{fig-1} compares $A_1(t)$ and $A_N(t)$ for $N=8$. In the next paragraph we will see that physically interesting quantities like current, fluctuations in current in NESS are related to the time integral of $C_{\Delta T}(t)$ and this integral depends only on $A_N(t)$, though $A_1(t)$ has dominant contribution in the correlation function itself. Hence it is more relevant to see the behaviour of $A_N(t)$ with system size $N$. In Fig.~\ref{fig-2} we plot $[A_N(t)]$ for different system sizes. Here we prefer to give plots of disordered averaged quantities, since very often we are interested in disorder averaged quantities. 

\begin{figure}
\includegraphics[width=9cm,height=5cm]{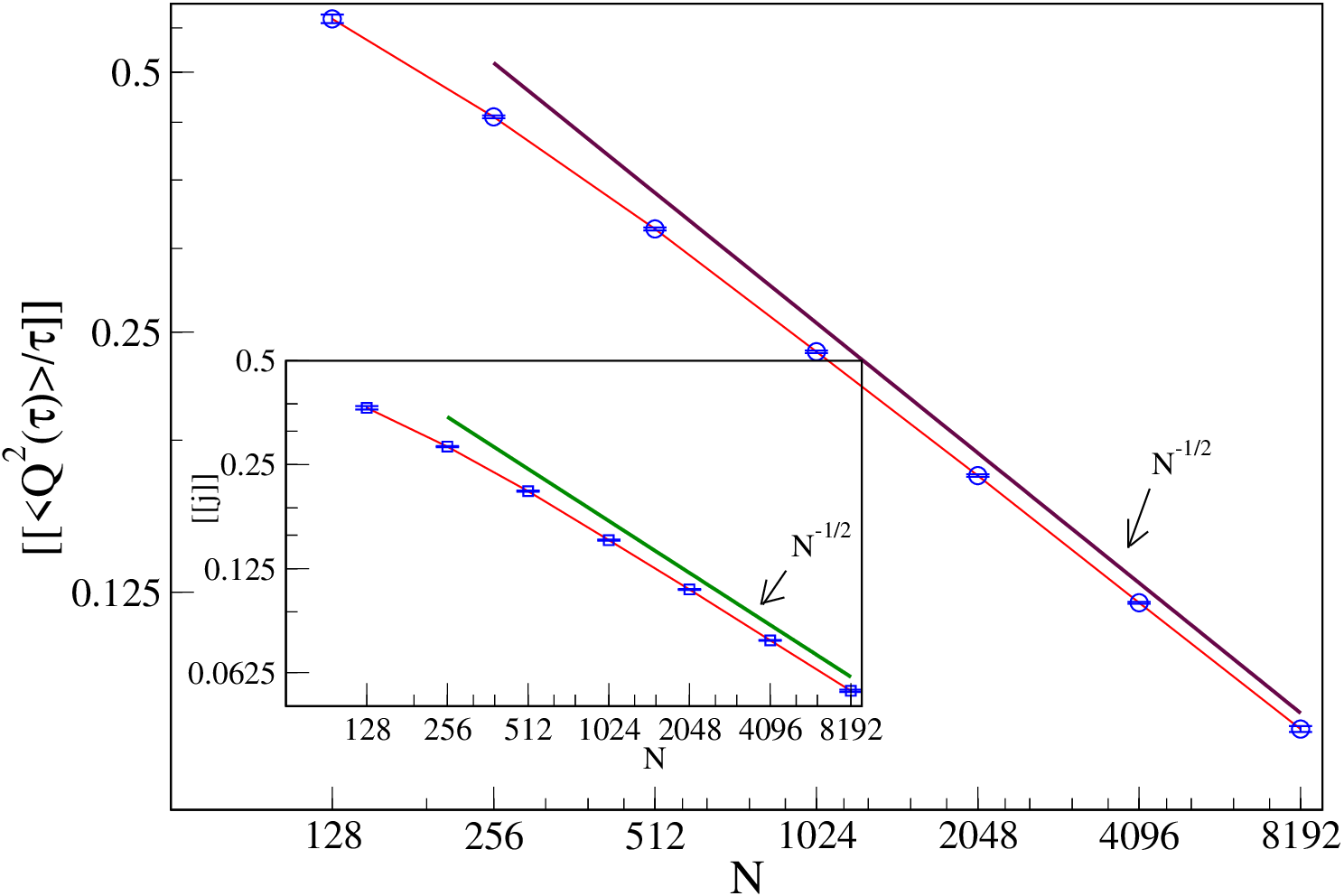}
\vspace{0.1cm}
\caption{(Color online) This figure shows the dependence of non-equilibrium current fluctuation on system size for free BC. The parameters for the figure are same as those for Fig.~\ref{fig-1} except $T_L=3.0$ and $T_R=2.0$. Inset shows the dependence of non-equilibrium current on system size for free BC. Disorder average is taken over 100 different disorder realizations. Standard deviation corresponding to each point is smaller that the size of the point symbol. 
}
\label{fig-3}
\end{figure}

Let $Q(\tau)=\int_o^{\tau}dt j_L(t)$ be the heat transfer in duration $\tau$ from left reservoir to the system. Using stationarity property of the correlation function it is easy to show that the $2^{nd}$ order cumulant of $Q(\tau)$ is related to 
$C_{\Delta T}(t)$ as
\bea
\lim_{\tau \to \infty} \f{\la Q^2(\tau) \ra_c}{\tau} = \int_{0}^{\infty} dt C_{\Delta T}(t).
\eea 
Now integrating the expression of $C_{\Delta T}(t)$ given in Eq.~(\ref{C_TLTR}) from $0$ to $\infty$ and again using the identity in Eq.({\ref{identity}}) we get
\bea
\int_{0}^{\infty} dt C_{\Delta T}(t) &=& 2\g_L\g_RT_LT_R A_N(0)\nn \\ 
&+& 8\g_L^2\g_R^2(T_L-T_R)^2\int_{0}^{\infty}dt~A_N^2(t)~. \nn \\
\label{int-corr}
\eea
In the frequency space the Eq.~(\ref{int-corr}) can be written as an integration over $\o$ of the transmission 
coefficient $\mathcal{T}(\o)$ defined in Eq.~(\ref{To}) and we obtain
\bea
\int_{0}^{\infty} dt C_{\Delta T}(t)&=&\f{(T_L-T_R)^2}{4\pi}\int_{0}^{\infty}d\o~\mathcal{T}^2(\o) \nn \\
&+&\f{T_LT_R}{2\pi}\int_{0}^{\infty}d\o~\mathcal{T}(\o)~. \label{int-corr-tomega}
\eea  
This expression matches with the expression given in \cite{keijidhar} for quantum mechanical systems,  
in the high temperature limit. Now if we put $T_L=T_R=T$ in the expression in Eq.~(\ref{int-corr}) and use Eq.({\ref{Jav}}) we get a relation between the current in the non-equilibrium steady state and the equilibrium correlation function similar to the GK relation derived in \cite{kundu}  
\bea
\int_{0}^{\infty} dt C_{0}(t) = \f{T^2}{2\pi}\int_{0}^{\infty}d\o~\mathcal{T}(\o)=T^2\f{j}{(T_L-T_R)},  
\eea
where $C_{0}(t)$ is the equilibrium auto-correlation function for the open system. The inset of Fig.~\ref{fig-3} shows the system size dependence of the disorder average of current. 

In general for large system sizes $[[j]]$ and $[\f{\la Q^2(\tau) \ra_c}{\tau}]$ scales with N as $N^{-\beta}$ and $N^{-\alpha}$ respectively. Using the frequency dependence of $T(\o)=[[\mathcal{T}(\omega)]]$  and $[[\mathcal{T}^2(\omega)]]$ one can predict the value of $\alpha$  and $\beta$ for different BC's.  
By computing $[[j]]$ in NESS, several authors have already studied asymptotic size dependence of $[[j]]$. Rubin and Greer \cite{Rubin} obtained $\beta=1/2$ for free BC, which was latter proved rigorously by Verheggen\cite{ver}. Casher and Lebowitz  \cite{lebo-cash} studied the same model and obtained a lower bound for $[[j]]\geq N^{-3/2}$ and simulations by Rich and vischer \cite{Rich} confirmed the exponent to be $\beta=3/2$. Later Dhar\cite{dhar01} obtained $j$ for both the boundary conditions using Langevin Equation and Green Function approach and obtained $\beta=1/2$ for free BC and $\beta=3/2$ for fixed BC. Here we follow the same procedure described in \cite{dhar01} to find the asymptotic size dependence of $[[\f{\la Q^2(\tau) \ra_c}{\tau}]]$ from the expression given in Eq.~(\ref{int-corr-tomega}).
 
We numerically observe that for both the BCs $[[\mathcal{T}^2(\omega)]]$ is much smaller than $T(\omega)$ for each $N$. 
Hence, in determining the assymptotic $N$ dependence, dominant contribution comes 
from the integration of $T(\omega)$ over $\o$. To determine $\alpha$, we use the fact (discussed in the first paragraph of this section) that for  $\omega$ greater than $\omega_d \sim N^{-1/2}$, $T(\omega)$ decays exponentially as $e^{-aN\omega^2}$ whereas, for $\omega < \omega_d$, $T(\omega)$ is almost identical to $\mathcal{T}_o(\omega)$ of an ordered chain. 
It can be shown that transmission coefficient of an ordered chain, 
denoted by $\mathcal{T}_o(\omega)$, is independent of $\omega$ for free BC and goes as $\omega^{2}$ for fixed BC. Now putting these forms of $\mathcal{T}_o(\omega)$ and integrating up-to $\omega_d \sim N^{-1/2}$ we get $\alpha=1/2$ 
for free BC and $3/2$ for fixed BC. We see that the asymptotic size dependence of current fluctuation is same as that of NESS current. We numerically evaluate the RHS of Eq.~(\ref{int-corr}) for free BC and obtain 
$\f{\la Q^2(\tau) \ra_c}{\tau}$ for $\tau \to \infty$ for different system sizes. In Fig.~\ref{fig-3} we plot $[[\f{\la Q^2(\tau) \ra_c}{\tau}]]$ versus system size $N$, which shows that the fluctuation in current scales 
with system size as $N^{-1/2}$, when both ends of the chain are free. In the pinned case, since there are no low frequency modes, $T(\o)$ decays exponentially and hence fluctuations in current decays exponentially with $N$.

\noindent
\section{ C\lowercase{onclusion}}\label{conclu}
In conclusion we have given an expression for the current-current correlation for a one dimensional mass-disordered harmonic system in NESS. The correlation function has been expressed in terms of the phonon Green's functions which are easy to evaluate numerically. We show that the integration of equilibrium correlation function gives current satisfying the finite size open system Green-Kubo formula whereas the integration of non-equilibrium correlation function gives information about current fluctuation in the NESS. Using the non-equilibrium correlation function we obtain asymptotic system size scaling of the fluctuation in the steady state current. A possible application of our results is that they can serve to test numerical codes for simulations studying correlations in non-equilibrium systems. In this paper we have considered a classical one dimensional system with white noise Langevin dynamics. It will be straightforward to get an expression for correlation function in quantum systems and higher dimensional systems.

\appendix
{
\section{Proof of Eq.[\ref{identity}]}
Let us first define few quantities:
\bea
\tilde{G}&=&M^{\f{1}{2}}GM^{\f{1}{2}} \nn \\
\tilde{\Gamma}&=&M^{-\f{1}{2}}\Gamma M^{-\f{1}{2}}\nn \\
\tilde{\Phi}&=&M^{-\f{1}{2}}\Phi M^{-\f{1}{2}} \nn 
\eea
Using this above definitions Eq.~(\ref{Geqm}) can be written as
\bea
\ddot{\tilde{G}}(t) + \tilde{\Gamma}\dot{\tilde{G}}(t) + \tilde{\Phi}\tilde{G}(t) = 0
\label{Gtildeeqm}
\eea
We use the above equation to evaluate $\f{d}{dt'}[\dot{\tilde{G^T}}(t')\dot{\tilde{G}}(t'+t)]$ and get
\bea
\f{d}{dt'}[\dot{\tilde{G^T}}(t')\dot{\tilde{G}}(t'+t)] &=&-2\dot{\tilde{G^T}}(t')\tilde{\Gamma}\dot{\tilde{G}}(t'+t) \nn \\ 
&+& \f{d}{dt'}[{\tilde{G^T}}(t')\tilde{\Phi}{\tilde{G}}(t'+t)] \nn
\eea
Now integrating both side of the above equation over $t'=0$ to $t'=\infty$ we get
\bea
\dot{\tilde{G}}(t)=  2 \int_{0}^{\infty}dt'\dot{\tilde{G^T}}(t')\tilde{\Gamma}\dot{\tilde{G}}(t'+t). \label{A4}
\eea
To the above equation we have used the following: $\dot{G}(0)=M^{-1}$, $G(0)=0,~G(t) \to 0$ as $t \to \infty$. Now we know that  $\Gamma_{ij}=(\f{\gamma_L}{m_1}\delta_{i1} + \f{\gamma_R}{m_N}\delta_{iN})\delta_{ij}$. Taking $(11)^{th}$ element on the both side of the matrix equation (\ref{A4}) we get
\bea
\f{\dot{G}_{11}(t)}{2} &=& \int_{0}^{\infty}dt'[\gamma_L\dot{G}_{11}(t')\dot{G}_{11}(t'+t) \nn \\ 
&+& \gamma_R\dot{G}_{1N}(t')\dot{G}_{1N}(t'+t)] \nn \\
&=& \gamma_LA_1(t) + \gamma_R A_N(t)
\eea
}
{
\section{Expressions of ${K_1}$'s}

\vspace{-1cm}
\bea
&&K_1^{(1)}(t_1,t_2,t_3,t_4,t)=\nn\\
&&\big[\g_L^2T_L^2\dot{G}_{11}(t-t_1)\dot{G}_{11}(t-t_2)\dot{G}_{11}(-t_3)\dot{G}_{11}(-t_4) \nn \\
&&+\g_R^2T_R^2\dot{G}_{1N}(t-t_1)\dot{G}_{1N}(t-t_2)\dot{G}_{1N}(-t_3)\dot{G}_{1N}(-t_4) \nn \\
&&+\g_LT_L\g_RT_R\big\{\dot{G}_{1N}(t-t_1)\dot{G}_{1N}(t-t_2)\dot{G}_{11}(-t_3)\dot{G}_{11}(-t_4) \nn \\  
&&+ \dot{G}_{11}(t-t_1)\dot{G}_{11}(t-t_2)\dot{G}_{1N}(-t_3)\dot{G}_{1N}(-t_4)\big\}\big], \nn 
\eea
----
\vspace{1cm}
\noindent 
\bea
&&K_1^{(2)}(t_1,t_2,t_3,t_4,t)=\nn\\
&&\big[\g_L^2T_L^2\dot{G}_{11}(t-t_1)\dot{G}_{11}(t-t_2)\dot{G}_{11}(-t_3)\dot{G}_{11}(-t_4) \nn \\
&&+\g_R^2T_R^2\dot{G}_{1N}(t-t_1)\dot{G}_{1N}(t-t_2)\dot{G}_{1N}(-t_3)\dot{G}_{1N}(-t_4) \nn \\
&&+\g_LT_L\g_RT_R\big\{\dot{G}_{1N}(t-t_1)\dot{G}_{11}(t-t_2)\dot{G}_{1N}(-t_3)\dot{G}_{11}(-t_4) \nn \\  
&&+ \dot{G}_{11}(t-t_1)\dot{G}_{1N}(t-t_2)\dot{G}_{11}(-t_3)\dot{G}_{1N}(-t_4)\big\}\big] \nn 
\eea 
and
\bea
&&K_1^{(3)}(t_1,t_2,t_3,t_4,t)=\nn\\
&&\big[\g_L^2T_L^2\dot{G}_{11}(t-t_1)\dot{G}_{11}(t-t_2)\dot{G}_{11}(-t_3)\dot{G}_{11}(-t_4) \nn \\
&&+\g_R^2T_R^2\dot{G}_{1N}(t-t_1)\dot{G}_{1N}(t-t_2)\dot{G}_{1N}(-t_3)\dot{G}_{1N}(-t_4) \nn \\
&&+\g_LT_L\g_RT_R\big\{\dot{G}_{11}(t-t_1)\dot{G}_{1N}(t-t_2)\dot{G}_{1N}(-t_3)\dot{G}_{11}(-t_4) \nn \\ 
&&+ \dot{G}_{1N}(t-t_1)\dot{G}_{11}(t-t_2)\dot{G}_{11}(-t_3)\dot{G}_{1N}(-t_4)\big\}\big] \nn 
\eea
}

\noindent
I thank Dr. Abhishek Dhar for useful suggestions and Jayakumar A for helpful discussions.

\end{document}